\documentclass[prb,twocolumn,showpacs,amsmath,amssymb,superscriptaddress,floatfix]{revtex4-1}
\usepackage{dsfont}

\usepackage{graphicx,color}
\usepackage{amsmath,amssymb,bm}

\begin{document}

\preprint{APS/123-QED}

\title{{\bf Fragility of Symmetry Protected Topological Order on a Hubbard Ladder}}

\author{Sanjay Moudgalya}
\affiliation{Indian Institute of Technology, Kanpur 208016, India}
\author{Frank Pollmann}
\affiliation{Max-Planck-Institut f\"{u}r Physik komplexer Systeme, 01187 Dresden, Germany\\
}

\begin{abstract}
Anfuso and Rosch [Phys. Rev. B {\bf 75}, 144420 (2007)] showed that the ``topological'' Haldane phase in a fermionic spin-$1/2$ ladder can be continuously deformed into a ``trivial'' phase without explicitly breaking symmetries when local charge fluctuations are taken into account.
Within the framework of symmetry protected topological phases, we revisit the model and demonstrate how the Haldane phase can be adiabatically connected to a trivial phase due to charge fluctuations.
Furthermore, we show that the Haldane phase remains stable as long as the system is symmetric under particular reflection symmetries.
\end{abstract}

\maketitle

\section{Introduction}
Different phases of matter are usually understood in terms of spontaneous symmetry breaking and can be detected by local order parameters.
For example, the $\mathds{Z}_2$ symmetric Ising model has a disordered (symmetric) and a ferromagnet (symmetry broken) phase which can be distinguished by measuring the magnetization as a local order parameter.
As long as the  $\mathds{Z}_2$  symmetry is not explicitly broken, it is not possible to adiabatically connect the two phases, i.e., there is necessarily a phase transition separating the two phases.  
Topological phases represent different kind of phases that are distinct from trivially disordered phases but cannot be characterized by symmetry breaking.
A particular class of topological phases are \emph{symmetry protected topological}  (SPT) phases, i.e., phases that are only distinct from a trivially disordered phase as long as certain symmetries are preserved. 
A well known example of an SPT phase is the  antiferromagnetic spin-$1$ Heisenberg chain.
As predicted by Haldane  \cite{Haldane-1983a}, an antiferromagnetic spin chain with integer spins has a gapped ground state with exponentially decaying correlations.
Following Haldane's prediction, Affleck, Kennedy, Lieb, and Tasaki (AKLT) presented model Hamiltonians  for which the ground state can be obtained exactly.\cite{Affleck87a}
In addition to providing insight to the Haldane conjecture, the AKLT state was later found to exhibit several interesting properties, such as a nonlocal ``string order'' and spin-$1/2$ edge states, which extend also to states within the same phase.\cite{denNijsRommelse} 
It was then realized that the Haldane phase can be understood in terms of ``symmetry fractionalization'', which is the defining property of SPT phases.  
That is, the degrees of freedom in the bulk of the system transforms linear under spin rotation symmetries (spin-1), while the edge transform projectively (spin-1/2).
The Haldane phase on the spin-1 chain has been shown to be protected any of the following symmetries: spatial inversion symmetry, time reversal symmetry or the $\mathds{Z}_2 \times \mathds{Z}_2 $ symmetry.\cite{Pollmann10}
More generally, different SPT phases are understood in terms of inequivalent projective representation of the symmetries on the edge degrees of freedom.
The projective representations are classified by the second cohomology group  $H^2[G,U(1)]$.\cite{ZCGu09,Pollmann10,XChen11a,XChen11b,Schuch-2011}
This classification scheme can then be applied to a wide range of one-dimensional models.\cite{Liu11d2h,Liu13optical,Motruk13,Tang12majorana,Duivenvoorden-2012,Nonne-2013}
Characteristic features of SPT phases include degeneracies of the entire \emph{entanglement spectrum} \cite{Li-2008} (i.e., the spectrum of the reduced density matrix) and the existence of non-local order parameters (e.g., the string order found in the AKLT state).\cite{Pollmann12b,Pollmann10}

However, SPT phases are rather fragile. 
Beside the symmetries that protect the phase, it is essential that the on-site representation of the symmetry is well defined.
In fact, before SPT phases were understood in terms the cohomology classification, Anfuso and Rosch \cite{Anfuso07}  constructed a fermionic two-leg ladder system that adiabatically connects the spin-1 Heisenberg point with a trivial product state by tuning it through a band-insulator phase. 
While their Hamiltonian explicitly breaks the inversion symmetry, the $\mathds{Z}_2 \times \mathds{Z}_2 $  and time reversal symmetries are preserved along the entire path in parameter space.
%
%

In this paper, we revisit the question about the fragility of SPT phases in the fermionic two-leg ladder using the framework of 1D SPT phases.
We first discuss how charge fluctuations can affect the SPT phase by a mixing of integer and half-integer on-site representations of the symmetry.
Using matrix-product state (MPS) based methods, we then calculate the entanglement spectrum along a path that adiabatically connects the Haldane  with the trivial phase.
From the entanglement spectrum, we observe a continuous crossing of the integer and half-integer part of the spectrum. 
Near the Heisenberg point, the low energy part consists of degenerate half-integer levels (which is the hallmark of an SPT phase), the high energy part of the spectrum contains integer levels (corresponding to a trivial phase). After passing through a crossover region, the spectrum is inverted.
For an unambiguous characterization of an SPT phase, \emph{all} states in the entanglement spectrum must belong to either half-integer or integer representations.\cite{Pollmann10}
A  related observation has been made in a different context  in a recent work by Chandran et al. Ref.~[\onlinecite{Chandran14}] that discusses the non-universality of the low-energy part of the entanglement spectrum.   
While the physical picture is nicely reflected in the evolution of the entanglement spectrum, we also show how charge fluctuations weaken the topological invariants of the SPT phase.
Finally, we show how the Haldane phase is protected in the presence of charge fluctuation by symmetry under combined reflections along the $x$- and $y$- axis of the ladder.

This paper is organized as follows.
We first introduce and discuss the model Hamiltonian used for our study in Section \ref{model}.
In Section \ref{SPTreview}, we briefly review SPT order in 1D and discuss the projective representations of the symmetry group that characterize the SPT phase.
In Section \ref{fragility}, we discuss the effect of charge fluctuations and its role in mixing symmetry representations.
In Section \ref{results}, we show our numerical results on the entanglement spectrum and the stability of the SPT phase in presence of reflection symmetries.
We conclude in Section \ref{conclusion} with implications of this effect to the universality of the entanglement spectrum and  SPT phases.

\section{Model}\label{model}
\begin{figure}
\includegraphics[scale = 0.45]{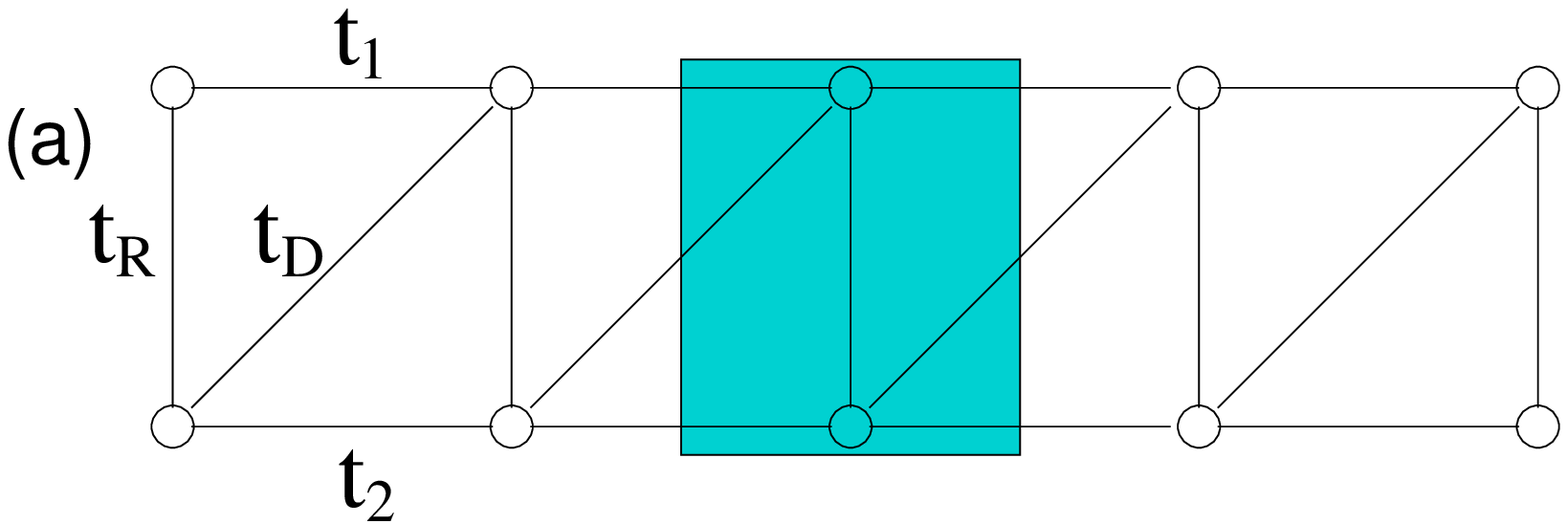}
\includegraphics[scale = 0.45]{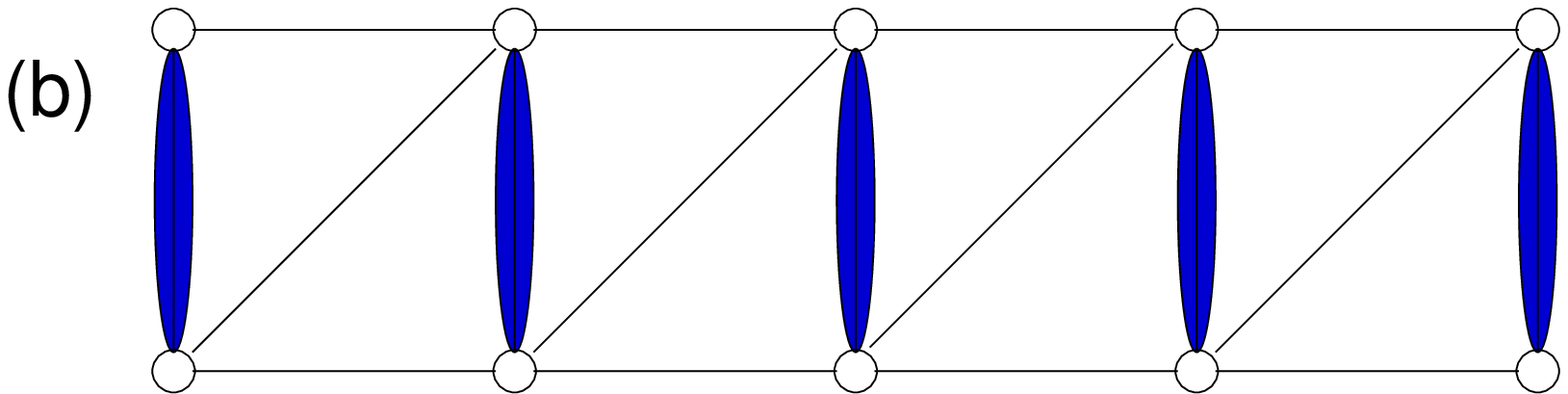}

\vspace{9pt}
\includegraphics[scale = 0.45]{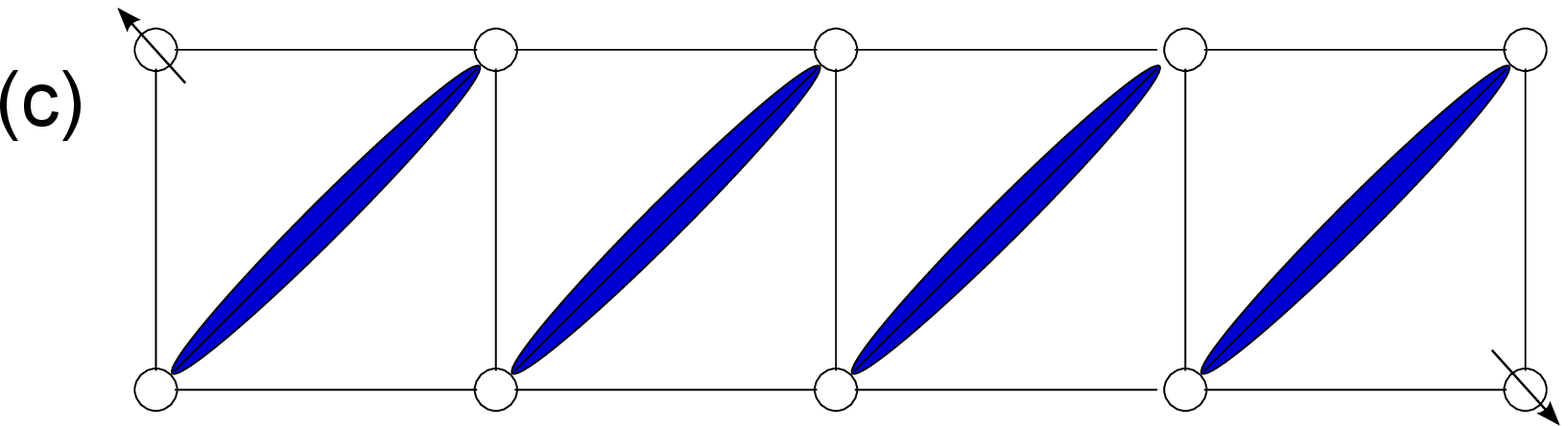}
\caption{(Color online) (a) Ladder structure with hopping amplitudes $t_1,t_2$ on the legs and $t_R$ and $t_D$ on the rungs and diagonals, respectively. The ladder has a two-site unit cell which is shaded in cyan. (b) Trivial product state of the spin ladder with rung singlets (shown by blue ellipsoids) representative for the ``rung-singlet phase''. (c) Non-trivial SPT state  with singlets on the diagonals and uncoupled edge spin-1/2 degrees of freedom representative for the ``bond-singlet phase''.\label{fig:latgeo}}
\end{figure}

We use a Hamiltonian describing a two-leg ladder on a lattice with the geometry as shown in Fig \ref{fig:latgeo}(a),  similar to the one used by  Anfuso et al. in Ref.~[\onlinecite{Anfuso07}]. 
Hopping terms are along the legs, rungs and one of the diagonals. 
Apart from that, there is an on-site potential $U$ punishing doubly occupied sites. 
The explicit form of the Hamiltonian is
\begin{equation}\label{orhamil}
\begin{split}
 H &= \sum_{i,\alpha,\sigma}{(-t_{\alpha}c^{\dagger}_{\alpha,i,\sigma}c_{\alpha,i+1,\sigma} + h.c. + \frac{U}{2}n_{\alpha,i,\sigma})}\\
 & + \sum_{i,\sigma}{(-t_{R} c^{\dagger}_{1,i,\sigma} c_{2,i,\sigma} - t_{D} c^{\dagger}_{1,i+1,\sigma} c_{2,i,\sigma})+ h.c.}\\ 
 &+ U \sum_{i,\alpha}{n_{\alpha,i,\uparrow} n_{\alpha,i,\downarrow}},
\end{split}
\end{equation}
where $\alpha$ counts the two legs of the ladder, $i$ represents sites along the ladder's length, and $\sigma$ counts the spins (up and down) on each site. 
This Hamiltonian spans a huge phase diagram with the parameters $\{t_{R},t_{D},t_{1},t_{2},U\}$.
To understand the phase diagram, we find it useful to first consider an effective low-energy model in the limit of large $U$ at half filling.
The Hamiltonian can then be mapped to a Heisenberg model by performing a Schrieffer-Wolff transformation \cite{Bravyi11} with two localized spin-1/2 degrees of freedom per unit cell:
\begin{equation}\label{swhamil}
\begin{split}
H &= \sum_{i}{\left(J_R\vec{S}_{1,i}.\vec{S}_{2,i} + J_D\vec{S}_{1,i+1}.\vec{S}_{2,i} + J_1\vec{S}_{1,i}.\vec{S}_{1,i+1}\right.}\\
&\left. + J_2\vec{S}_{2,i}.\vec{S}_{2,i+1}\right),
\end{split}
\end{equation}
where $J_i = 4 t_i^2/U$. 
The phase diagram of this ladder can be easily visualized by unfolding it into a 1D chain by considering the sites along the rungs and the diagonals as nearest neighbors.
Consequently, the 1D spin chain has alternating interaction strengths $J_D$ and $J_R$ between nearest neighbors and $J_1$ and $J_2$ between alternating next-nearest neighbors.
If  $J_D < J_R$ and $J_1 = J_2 = 0$, the ground state is a unique state of singlets between one set of nearest neighbors (rungs) which is protected by a finite gap (the energy needed to break one singlet).
This state is adiabatically connected to a  ``rung-singlet'' products state shown in Fig \ref{fig:latgeo}(b).
If  $J_D > J_R$ and $J_1 = J_2 = 0$, the ground state is again gapped but now forms singlets between the other set of nearest neighbors (diagonals).
This state is adiabatically connected to a  ``bond-singlet''  state shown in Fig \ref{fig:latgeo}(c).
The two free spin-1/2 degrees of freedom at the edges yield a robust  four fold degeneracy.
In fact, this state can be adiabatically transformed into a {spin-1} Haldane chain by projecting each rung onto an $S=1$ state (e.g., by introducing a ferromagnetic rung-coupling). \cite{EHKim00}
At $J_D = J_R$, the gap closes and system undergoes a phase transition between the rung-singlet and the bond-singlet phase.
Moreover, it can be shown that an adiabatic connection of the two phases is still not possible by varying $J_1$ and $J_2$. \cite{Watanabe99,Shelton96}
As emphasized by Bonesteel et al. in Ref.~[\onlinecite{Bonesteel89}], the phases for $J_R < J_D$ and $J_R > J_D$ are distinct by the fact that cutting a vertical line in the ground state wavefunction would cut an odd number of singlets in the bond-singlet phase and an even number of them in the rung-singlet phase.
This is in fact directly related to the distinction of different SPT phases as discussed in the following section.

Let us now consider the full Hamiltonian Eq.~(\ref{orhamil}).
In Ref.~[\onlinecite{Anfuso07}], Anfuso and Rosch showed that a path adiabatically connecting the bond-singlet (topological) and rung-singlet (trivial) phase exists in the parameter space.
The path is shown in Fig.~\ref{fig:phase}.
In particular, the path found connects the two ground states with $t_D=1, t_R=0$ and $t_D=0, t_R=1$ at $ t_1=t_2=0$, $U\rightarrow\infty$ (i.e., the rung- and the bond-singlet states). 
For this, the interaction strength is first adiabatically lowered to $U=0$ while keeping  $t_D=1, t_R=0$ and  $t_1=t_2=0$ fixed.
The non-interacting fermion system is then solved exactly for which the two bands have energies \cite{Anfuso07}
\begin{equation}
\begin{split}
E\left(k\right) &= \left(t_{1} + t_{2}\right)\cos\left(k a\right)\pm \{\left(t_{1} - t_{2}\right)^{2} \cos^{2}\left(k a\right)\\ 
&+ \left(t_{R} - t_{D}\right)^{2} + 2 t_{R} t_{D} \left[1 + \cos\left(k a\right)\right]\}^{\frac{1}{2}}.
\end{split}
\end{equation}
It is easy to see that a path connecting the two limits $t_R = 1,t_D=0$ and  $t_R = 0,t_D=1$ without closing the gap can be found for some $t_1 \ne t_2$.  
Thereafter, the interaction strength is again increased adiabatically to the limit of $U\rightarrow\infty$ while keeping $t_D=0, t_R=1$ and  $t_1=t_2=0$ fixed.
By this, the two phases, that were distinct SPT phases in the Heisenberg model Eq.~(\ref{swhamil}), are now adiabatically connected without breaking the symmetries protecting it (e.g., the spin rotation or time reversal symmetry)!

\section{Brief review of SPT order in 1D} \label{SPTreview}
Here we briefly reiterate the concept of SPT phases in 1D and introduce the notations used.
For a complete discussion, we refer to the existing literature on this subject. \cite{Pollmann10,Pollmann12a,Schuch-2011,XChen11a}
Ground states of gapped 1D local Hamiltonians can be efficiently represented by a Matrix Product State (MPS).\cite{Verstraete-2006}
The MPS of a translationally invariant, infinite chain can be expressed in the canonical form \cite{Vidal-2007}
\begin{equation}
|\psi\rangle = \sum_{\left\{j\right\}}{\text{Tr}\left[\dots\Gamma^{j_1}\Lambda\Gamma^{j_2}\Lambda \dots\right]|\dots j_1 j_2 \dots\rangle}.\label{eq:mps}
\end{equation}
Here $\Gamma^{j_i}$ is a $d \times \chi \times \chi$ tensor where $d$ is the dimension of the local Hilbert space at lattice site $i$.
The matrices are chosen such that multiplying all matrices to the left (right) of a given bond yield left (right) Schmidt states of a Schmidt decomposition of the chain into two half chains at that bond.
The $\chi \times \chi$ diagonal matrix $\Lambda$  on the bond contains the corresponding Schmidt values.
Recall that in the \emph{Schmidt decomposition}, a state $|\Psi\rangle\in\mathcal{H}$ is decomposed as
\begin{equation}
|\Psi\rangle = \sum_{\alpha} \Lambda_\alpha |\alpha\rangle_L \otimes |\alpha\rangle_R, \quad |\alpha\rangle_{L(R)} \in \mathcal{H}_{L(R)} ,\label{eq:schmidt}
\end{equation}
where the states $\{|\alpha\rangle_{L(R)}\}$ form an orthogonal basis of Hilbert space describing the left (right) part of the cut $\mathcal{H}_L$ ($ \mathcal{H}_R$) and  $\Lambda_\alpha \ge 0$.
The \emph{entanglement spectrum} $\{\epsilon_\alpha\}$ \cite{Li-2008} is defined in terms of  the Schmidt spectrum $\{\Lambda_{\alpha}\}$  by
\begin{equation}
\Lambda_\alpha^2 = \exp(-\epsilon_\alpha)\label{eq:es}
\end{equation}
for each $\alpha$.
If an MPS in canonical form is invariant under an \emph{internal symmetry operation} $g\in G$ represented in the spin basis as a unitary matrix $u_g$ that is applied to all sites, then the $\Gamma^j $ matrices must transform under  $u_g$  in such a way that the product in Eq.~(\ref{eq:mps}) does not change (up to a phase). 
The transformed matrices can thus be shown to satisfy \cite{Pollmann10,PerezGarcia08}
\begin{equation}
\sum_{j^{\prime }}(u_g) _{jj^{\prime }}\Gamma ^{j^{\prime }}=e^{i\theta_g}U_g^{\dagger }\Gamma ^{j}U^{\vphantom{\dagger }}_g\text{,}  \label{trans}
\end{equation}
where $U_g$ is a unitary matrix that commutes with the $\Lambda $
matrices, and $e^{i\theta_g}$ is a phase factor.
As the symmetry element $g$ is varied over the whole group, a set of phases $e^{i\theta_g}$ and matrices $U_g$ results. 
The phases form a 1D representation (i.e., a character) of the symmetry group.  
The matrices $U_g$ form a $\chi-$dimensional (projective) representation of the symmetry group. 
A projective representation is an ordinary regular representation but up to phase factors.
For example, if $gh=k$ with $g,h\in G$, then  $U_g U_h=e^{i\rho(g,h)}U_k$. 
The phase angles $\rho(g,h)$ are called the ``factor set" of the representation.
If all phase factors are identities, the representation is a linear one. 
This factor set can be used to classify different  symmetry protected topological phases in 1D. \cite{Pollmann10,Pollmann12a,Schuch-2011,XChen11a}
Consider for example a model with localized integer degrees of freedom on each site that is invariant under a $\mathbb{Z}_2\times \mathbb{Z}_2$ symmetry of rotations $\mathcal{R}_x=\exp(i\pi S^x)$ and $\mathcal{R}_z=\exp(i\pi S^z)$. 
The phases for each spin rotation (e.g., $U_{x}^2=e^{i\alpha}\mathds{1}$) can individually be removed by redefining the phase of the corresponding $U$-matrix.  
However, the representations of  $\mathcal{R}_x\mathcal{R}_z$ and $\mathcal{R}_z\mathcal{R}_x$ can also differ by a phase, which turns out must be $\pm 1$.
Thus there are two different SPT phases in the presence of $\mathbb{Z}_2\times \mathbb{Z}_2$ which differ by the  the gauge invariant quantity  
\begin{equation}
P = U^{\vphantom{\dagger}}_x U^{\vphantom{\dagger}}_z U_x^{\dagger} U_z^{\dagger}.
\label{eq:P}
\end{equation}
If $P=\mathds{1}$, the state is in a trivial phase and can be adiabatically connected to a product state while preserving the $\mathbb{Z}_2\times \mathbb{Z}_2$ symmetry, while it is in an SPT phase that cannot adiabatically be a product state if $P=-\mathds{1}$. 
As shown in Ref.~[\onlinecite{Pollmann10}], the latter case implies that the entire even degeneracy must to have an even degeneracy.

This concept can now be directly applied to the effective Heisenberg ladder Eq.~(\ref{swhamil}).
For this we first  group the two-site unit cell to one site with local dimension $d=4$ so that the resulting model has an integer spin per site (i.e., S=0,1).
A state with perfect rung-singlet order is a simple product state for which we find that $U_x=U_z=\mathds{1}$ and thus $P=\mathds{1}$. 
A bond-singlet state is an MPS of bond dimension $\chi=2$ that transforms under the spin rotations by $U_x = \sigma_x$ and  $U_z= \sigma_z$ with $\sigma_x,\sigma_z$ being the Pauli matrices.
Since the Pauli matrices anti-commute, we find that $P=\mathds{-1}$ and thus the phase is an SPT phase, actually the Haldane phase. 

\section{Fragility of SPT order} \label{fragility}
The full Hamiltonian Eq.~(\ref{orhamil}) allows the fermions to hop and thus one can have either integer or half-integer per unit-cell, i.e., the total spin per unit cell is $S=0$, $1/2$, or $1$.
Let us again group the two-site unit cell to one site which has a local dimension of $d=16$.
The on-site representation of the spin-rotations is  either linear or projective, depending on whether an even or odd number of fermions present.
This mixing of the representation leads to a so-called ``grading'' of the representation.
It is easy to see that the $\pi$-rotation operators $\mathcal{R}_x$ and $\mathcal{R}_z$ acting on a single spin-1/2 fermion are given by the corresponding Paul matrices.
On a state with an even (odd) number of fermions, they are just the product of even (odd) number of Pauli matrices. 
Since the rotation operators $\mathcal{R}_x$ and $\mathcal{R}_z$ do not change the spin of the sites they act on and products of an even (odd) number of Pauli matrices commute (anti-commute), they must be $\mathbb{Z}_2$-graded.
That is, they split into two parts that act separately on the integer and half-integer states without mixing them.
This is also evident from the fact that the algebra $\mathcal{C}$ of the Pauli matrices (Clifford algebra) is $\mathbb{Z}_2$-graded.
That is, the operators in the algebra split into two parts that do not mix with operators containing even ($\mathcal{C}^0$) and odd number ($\mathcal{C}^1$) of Pauli matrices and $\mathcal{C} = \mathcal{C}^0 \oplus \mathcal{C}^1$. \cite{Garling11}
Because of the properties of $\mathbb{Z}_2$-graded algebras, operators in $\mathcal{C}^0$ commute whereas the operators in $\mathcal{C}^1$ anti-commute.
Hence any representation of the rotation operators operators (including the $U$ matrices) are block-diagonalized into integer and half-integer parts.
The matrix $P$, as defined in Eq.~(\ref{eq:P}), thus has the form $P = (+\mathds{1} )\oplus (-\mathds{1})$.
%
%
%
%
Thus, $P$ is a $\chi$ dimensional diagonal matrix with $p_{\alpha}=\pm 1$ along its diagonal, depending on whether the corresponding basis state transforms linearly or projectively.
Since $P$ commutes with the matrix $\Lambda$, we can associate a phase $p_\alpha$ with each Schmidt value $\Lambda_{\alpha}$. 
As argued above for the Haldane phase, every $\Lambda_{\alpha}$ corresponding to $p_\alpha = -1$ has an even degeneracy. 
The charge fluctuations present at any finite $U<\infty$  causes the non-zero Schmidt values  of the ground state to contribute to both sectors.
Starting from the Haldane phase, we can then for example continuously increase the Schmidt values for the integer states, and correspondingly decrease the Schmidt values for the half-integer states.
This can be continuously taken to the limit where the Schmidt values for the half-integer states vanish, and all the states are integer states.
Consequently, the ``fractionalized edge modes'' can mix with the bulk and there is no topological separation between the trivial phase and the Haldane phase when charge fluctuations are allowed. 
A similar argument can be made for the time-reversal symmetry that protects the SPT order in the absence of charge-fluctuations and the same fragility applies.
Note that charge fluctuations are absent in Eq.~(\ref{swhamil}) (after a Schrieffer-Wolff transformation), and the SPT phase is well defined.  
\begin{figure}
\includegraphics[scale = 0.5]{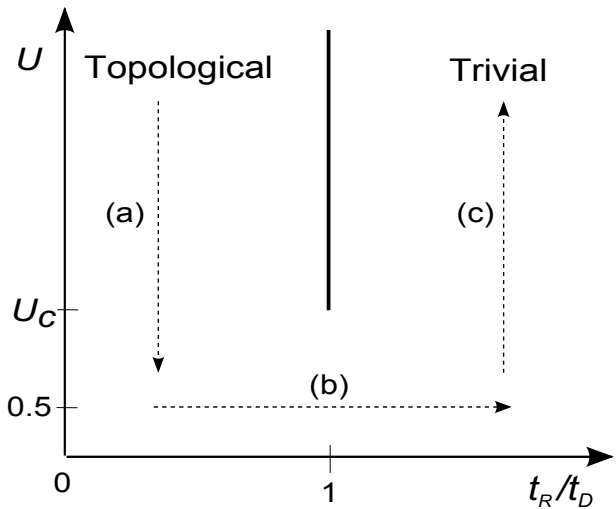}
\caption{The phase diagram for Hamiltonian Eq.~(\ref{orhamil}) for $t_1 \ne t_2$. The dashed lines indicate the path for adiabatic connection of the two phases. (a) From the ``topological" bond-singlet phase, $U$ can be continuously decreased from a large value to a small value below $U_c$. (b) For any value $U < U_c$, $t_R$ and $t_D$ can be tuned from $t_R/t_D < 1$ to $t_R/t_D > 1$ without a phase transition. (c) $U$ can then be continuously increased to a large value, leading to the ``trivial" rung-singlet phase.}
\label{fig:phase}
\end{figure}
\begin{figure*}[ht]
 \begin{tabular}{cc}
\includegraphics[scale = 0.43]{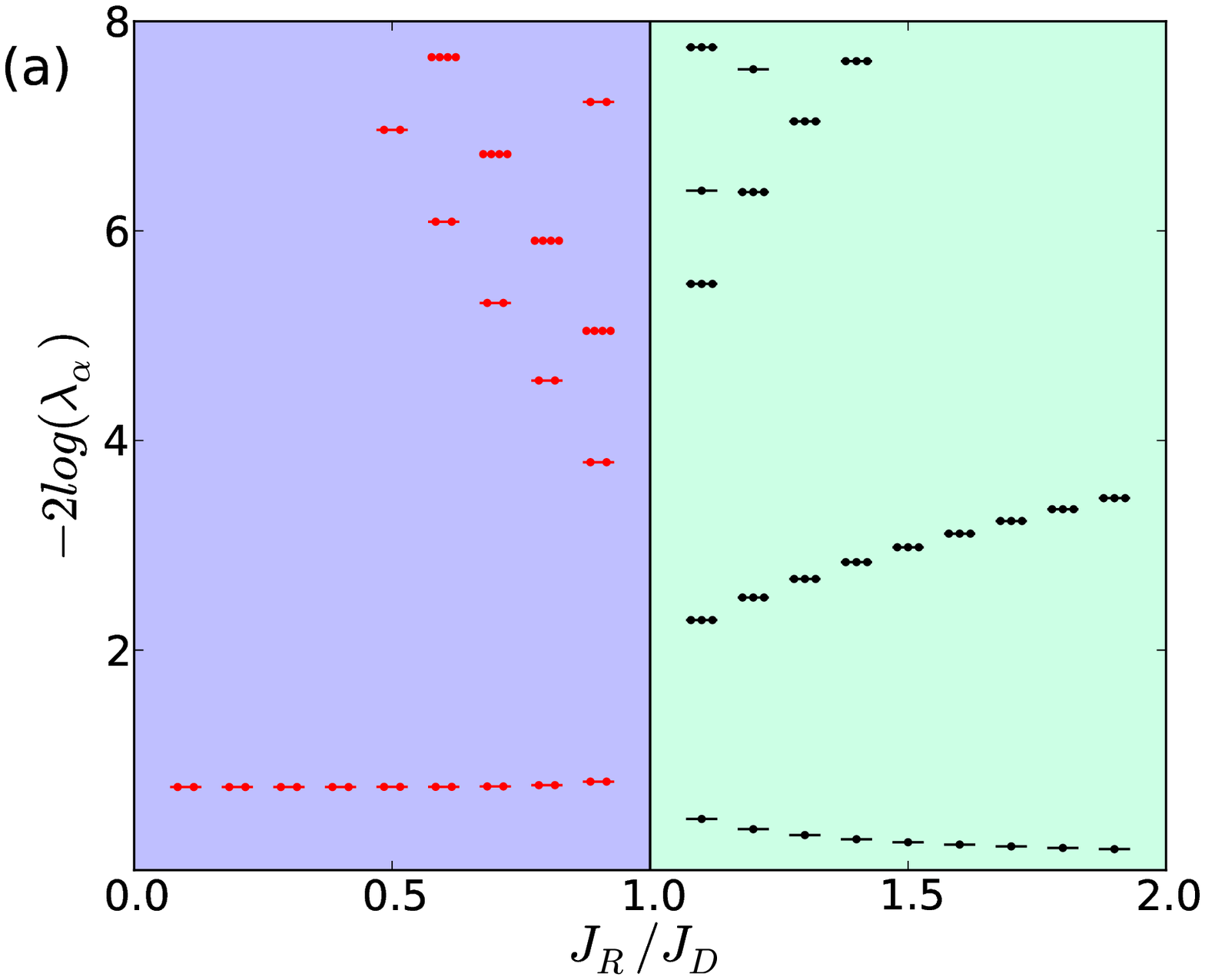}&\includegraphics[scale = 0.43]{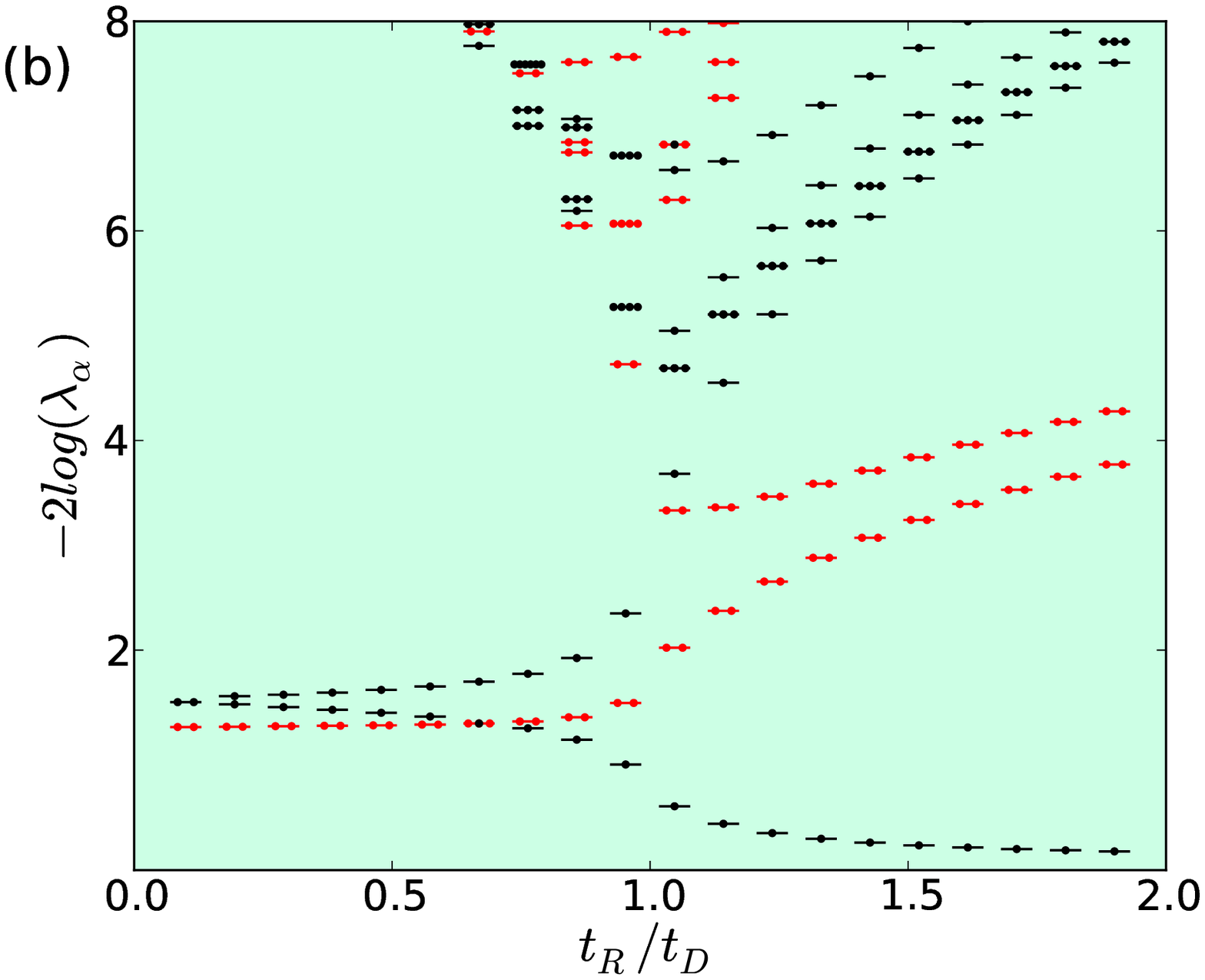}\\
\includegraphics[scale = 0.43]{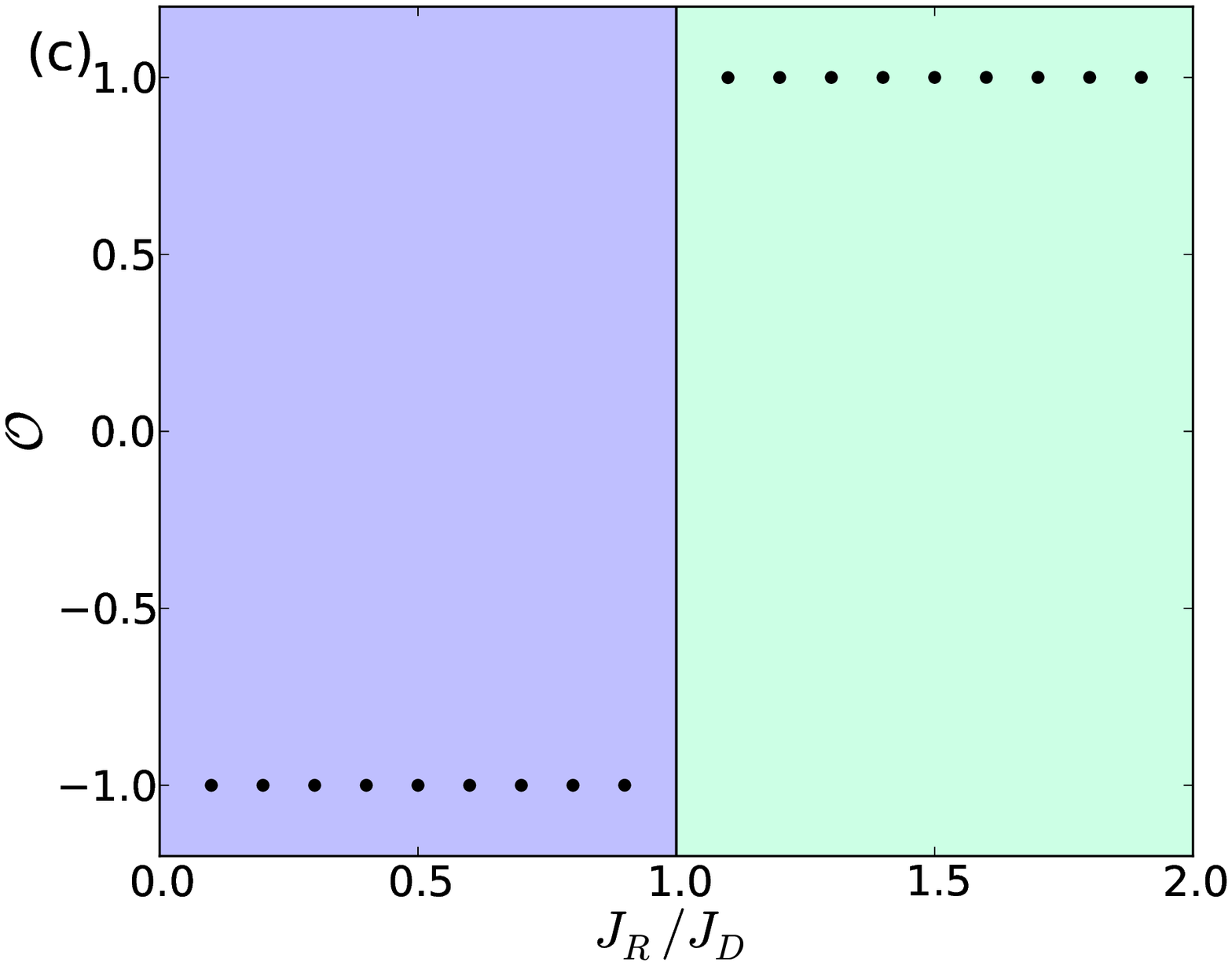}&\includegraphics[scale = 0.43]{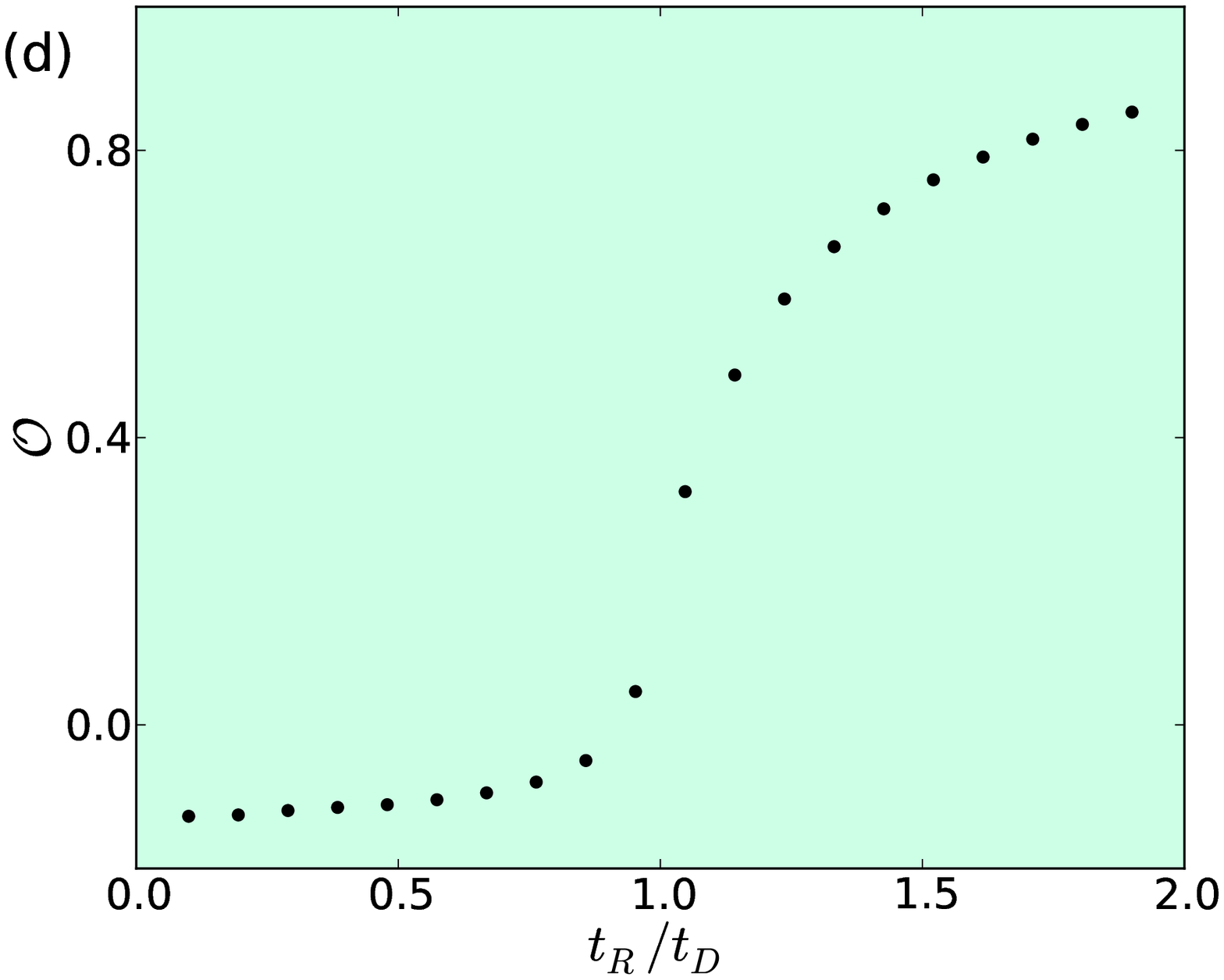}\\
\end{tabular}
\caption{(Color online) The entanglement spectrum for (a) the effective Heisenberg model Eq.~(\ref{swhamil}) and (b) the full model  Eq.~(\ref{orhamil}). The color of the dots indicates whether the corresponding Schmidt states yield half-integer (red) or integer (black) representation of the spin rotation symmetries. The lower panels (c) and (d) show the evolution of the corresponding ``non-local'' order parameter for the two models (see text for details). \label{fig:esord}}
\end{figure*}

\section{Numerical Results}\label{results}

We now study Hamiltonians Eqs.~(\ref{orhamil}) and (\ref{swhamil}) while tuning from the Haldane phase to the trivial phase by varying $t_R/t_D$.
The ground state is found using the infinite Time Evolution Block Decimation (iTEBD) algorithm. \cite{Vidal-2007}
For our simulations, we make use of  two $U(1)$ symmetries, conserving both magnetization and total charge in the system.
The main results are summarized in Fig \ref{fig:esord}.

\subsection{Without Reflection symmetry ($t_1 \ne t_2$)}
We first consider the less symmetric case with $t_1 \ne t_2$ for which the Haldane phase is protected by spin rotation and time reversal symmetry.
Figure \ref{fig:esord}(a) shows the even degeneracy of the effective Heisenberg model Eq.~(\ref{orhamil}) for $J_1 = 0.01$ and $J_2 = 0.04$.
We observe a phase transition at $J_D=J_R$ from the Haldane phase, in which \emph{all} states in the entanglement spectrum  have an even degeneracy, to a trivial phase with only accidental degeneracies. 
The states represented in red transform projectively under the symmetry (half-integer spin with $p_\alpha = -1$) and the state represented in black transform linearly (integer spin with $p_\alpha = +1$).
At the point of the phase transition, the correlation length diverges and the ``non-local'' order parameter \cite{Pollmann-2012} 
\begin{equation}\label{ord}
 \mathcal{O} = Tr\left(U^{\vphantom{\dagger}}_{x}U^{\vphantom{\dagger}}_{z}U_{x}^{\dagger}U_{z}^{\dagger}\right)/\chi
\end{equation}
shows a jump from $-1$ to $+1$ in Fig.~\ref{fig:esord}.
Fig.~\ref{fig:phase} shows the phase diagram for the full Hamiltonian Eq.~(\ref{orhamil}).
The entanglement spectrum as a function of $t_{R}/t_{D}$ for $U = 0.5$, $t_{1} = 0.1$ and $t_{2} = 0.2$ (path Fig.~\ref{fig:phase}(b)) is plotted in Fig.~\ref{fig:esord}(b).
%
%
Most noticeably, no phase transition occurs and no singularities are visible along the path.
In the topological phase in the upper left corner of Fig.~\ref{fig:phase}, the integer states are high up in the entanglement spectrum, and the entire spectrum consists of half-integer states.
Along path Fig.~\ref{fig:phase}(a), the integer states descend in the entanglement spectrum.
In Fig.~\ref{fig:esord}(b), for small $t_R/t_D$, the upper and lower doublets correspond to integer and half-integer states respectively.
As one increases the rung hopping $t_R$, the half-integer levels (red) remain degenerate because they are protected by the projective nature of the symmetry whereas the integer levels (black) split.
In the low energy entanglement spectrum, there is a crossover between the integer levels and the protected half-integer doublets.
This is a manifestation of the Schmidt values corresponding to integer states exceeding the Schmidt values for the half-integer states due to the grading of the representations.
This effect also becomes clear in the order parameter defined in Eq.~(\ref{ord}) and plotted in Fig.~\ref{fig:esord}(d).
It continuously changes from negative to positive values, indicating that the phases can be adiabatically connected.
An adiabatic crossover is found for any value of $U < U_c$ due to the mixing of inequivalent representations of the $\mathbb{Z}_2 \times \mathbb{Z}_2$ symmetry.
The critical $U_c$ is numerical very difficult to determine using the iTEBD method due to the large entanglement -- from simulations with bond dimensions $\chi\le150$ we estimate $U_c\approx1.7$ for the hopping amplitudes as defined above.
The reason for expecting the gap to close for $U>U_c$ is due to the similarity of this model with the 1D ionic Hubbard model as discussed in Ref.~[\onlinecite{Anfuso07}].
This implies in terms of the entanglement spectrum that the integer states still split but the Schmidt values do not exceed those for the half-integer states before $t_R/t_D = 1$.
Note that if $t_R/t_D$ is tuned for a large value of $U > U_c$  along the path Fig.~\ref{fig:phase}(b), the low energy part of the spectrum would look identical to Fig.~\ref{fig:esord}(a).
\subsection{With reflection symmetry ($t_1 = t_2$)}
\begin{figure}
\includegraphics[scale = 0.41]{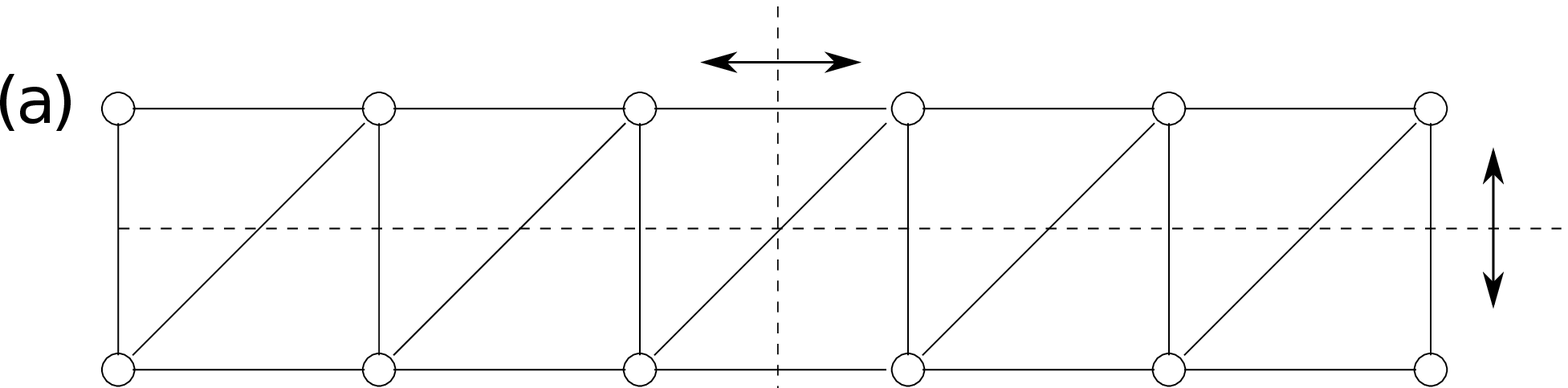}

\vspace{10pt}
\includegraphics[scale = 0.43]{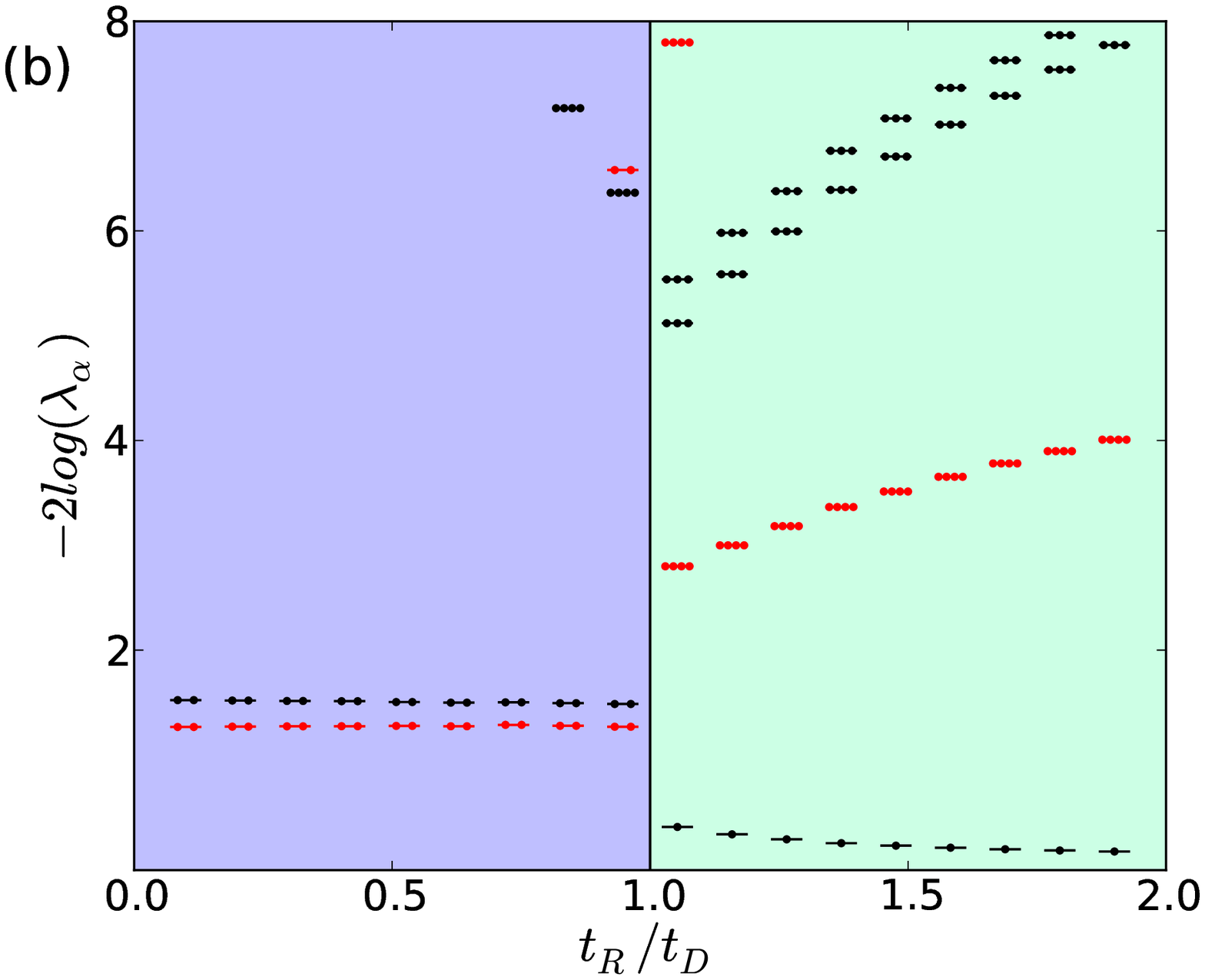}
\caption{(Color online) The reflection symmetric case of Hamiltonian Eq. (\ref{orhamil}) with $t_1 = t_2$ (a) The operation on the lattice that leaves the Hamiltonian Eq.~(\ref{orhamil}) and the state invariant. The dashed lines represent reflection about the line. (b) Entanglement Spectrum for $t_1 = t_2$. The color of the dots indicates whether the corresponding Schmidt states yield half-integer (red) or integer (black) representation of spin rotation symmetries. }
\label{fig:reflection}
\end{figure}

In the presence of additional (spacial) symmetries, the Haldane phase turns out to be robust.  
If $t_1 = t_2$ in Eq.~(\ref{orhamil}), one cannot avoid the phase transition, even with charge fluctuations.
The entanglement spectrum for this case with $t_1 = t_2 = 0$ is plotted in Fig \ref{fig:reflection}(b).
There is a phase transition at $t_D = t_R$ and as seen in the entanglement spectrum, the $t_R / t_D < 1$ phase has an even degeneracy everywhere, suggesting the existence of a symmetry protected topological (SPT) phase.

Setting $t_1 = t_2$ introduces a combined reflection symmetry of the lattice, where the ladder is inverted about the vertical axis, and sites on the top and bottom chains are interchanged, as shown in Fig \ref{fig:reflection}. 
To check if this is the symmetry that protects the phase, we look at the projective representations in the edge.
Correspondingly, in the MPS, $\Gamma \rightarrow \Gamma^T$ (because of the inversion symmetry about the vertical axis) combined with an on-site operator $f$ that flips the top and bottom sites in the unit cell.
The equation for transformation of the MPS is
\begin{equation}
\sum_{j^{\prime }}f_{jj^{\prime }}\Gamma^T _{j^{\prime }}=e^{i\theta_R}U_R^{\dagger }\Gamma _{j}U^{\vphantom{\dagger }}_R 
\end{equation}
The  representations $U_R$ for this symmetry operation can be calculated as prescribed by Ref.~[\onlinecite{Pollmann10}], and one notices that this symmetry indeed protects the Haldane phase. 
The order parameter corresponding to the combined reflection symmetry \cite{Pollmann-2012} can be used to distinguish the two phases by
\begin{equation}
\mathcal{O} = Tr(U^{\vphantom{*}}_R U^{*}_{R})/\chi.
\end{equation}
This quantity takes the value of $-1$ in the Haldane phase and $+1$ in the trivial phase. 
As seen in the entanglement spectrum in Fig \ref{fig:reflection}(b), the $\mathbb{Z}_2 \times \mathbb{Z}_2$ symmetry is still graded and does not protect the phase.

\section{Conclusion}\label{conclusion}

We revisited the question about the fragility of the topological Haldane phases in a fermionic Hubbard ladder originally proposed by Anfuso and Rosch [\onlinecite{Anfuso07}] in the framework of symmetry protected phases.   
We showed that  charge fluctuations allow for an adiabatic path between the ``topological'' Haldane phase and a ``trivial'' phase while preserving symmetries as integer and half-integer on-site representations of the symmetry group are mixed.
Using matrix-product state (MPS) based methods, we numerically calculated the entanglement spectra along a path connecting the two phases and observed a continuous crossing of the integer and half-integer part of the spectrum. 
If the charge fluctuations are projected out (Schrieffer-Wolff transformations), we found  that the Haldane phase is well defined and separated from the trivial phase through a phase transition. 
Furthermore, we showed that the Haldane phase is protected in the presence of charge fluctuation in the presence of a combined reflection symmetry of the ladder. 
On a different note, Chandran et al. [\onlinecite{Chandran14}] have questioned the validity of the universal nature of the low-energy part of the entanglement spectrum. 
They argued that physical observables are obtained at an entanglement temperature of $T_{E} = 1$ in the Entanglement Hamiltonian $H_E$ defined as $\rho_{red} = \exp{\left(-H_{E}/T_E\right)}$, whereas the low energy entanglement spectrum corresponds to $T_{E} \rightarrow 0$. 
They also illustrated in models where $H_{E}$ can undergo a phase transition when $H$ does not, meaning a phase transition in the boundary but not in the bulk. 
In the model considered here, if one simply looks at the low energy entanglement spectrum to characterize the phase in Fig \ref{fig:esord}(b), one would infer that there are two distinct phases.
However, they are the same phase as the two limits are connected by a smooth crossover.
It is thus essential that the entire spectrum shares the same properties for a phase to be well defined.

\section{Acknowledgement}
We thank Ari Turner and Karlo Penc for stimulating discussion and comments on the manuscript. 
SM thanks the Deutscher Akademischer Austausch Dienst (DAAD) for providing the financial support for this work. SM also acknowledges the hospitality of Visitors Program, MPIPKS.

\bibliography{bibo}

\end{document}